\documentclass[12pt]{iopart}

\usepackage{graphicx}
\usepackage[hidelinks]{hyperref}

\usepackage[percent]{overpic} % text on top of graphics
\usepackage{xcolor} % to mark text
\usepackage{upgreek}
\usepackage{units}
\usepackage{blindtext}

\newcommand{\edd}{\varepsilon_\mathrm{dd}}
\newcommand{\as}{a_\mathrm{s}}
\newcommand{\add}{a_\mathrm{dd}}

\renewcommand{\vec}[1]{\mathbf{#1}}

\newcommand{\textold}[1]{\textcolor{red!90!black}{#1}}
\newcommand{\textnew}[1]{\textcolor{green!50!black}{#1}}

\renewcommand{\textold}[1]{}
\renewcommand{\textnew}[1]{#1}

\begin{document}

\title[Quantum droplets and dipolar supersolid]{New states of matter with fine-tuned interactions: quantum droplets and dipolar supersolids}

%\author{Dysi Team$^1$}
\author{Fabian B\"{o}ttcher$^{1, \footnotemark[1]}$, Jan-Niklas Schmidt$^1$, Jens Hertkorn$^1$, Kevin S. H. Ng$^1$, Sean D. Graham$^1$, Mingyang Guo$^1$, Tim Langen$^{1, \footnotemark[1]}$ and Tilman Pfau$^1$ \footnotetext[1]{These authors contributed equally to this work.}}
\ead{t.pfau@physik.uni-stuttgart.de}
\address{$^1$ 5{.} Physikalisches Institut and Center for Integrated Quantum Science and Technology, Universit{\"a}t Stuttgart, Pfaffenwaldring 57, 70569 Stuttgart, Germany}
\date{\today}

\begin{abstract}\\%% ~200 words
%Quantum fluctuations can stabilize Bose-Einstein condensates with two types of interactions against a mean-field collapse. In this process, the fine-tuning of these interactions can lead to the emergence of two novel states of matter, liquid-like self-bound quantum droplets and supersolid crystals formed from these droplets. Here, we present an overview of the properties of these exotic states of matter, summarize the rapid experimental progress in their study using dipolar quantum gases and Bose-Bose mixtures and outline important open questions that need to be addressed in the future. 
Quantum fluctuations can stabilize Bose-Einstein condensates (BEC) against the mean-field collapse. Stabilization of the condensate has been observed in quantum degenerate Bose-Bose mixtures and dipolar BECs. The fine-tuning of the interatomic interactions can lead to the emergence of two new states of matter: liquid-like self-bound quantum droplets and supersolid crystals formed from these droplets. We review the properties of these exotic states of matter and summarize the experimental progress made using dipolar quantum gases and Bose-Bose mixtures. We conclude with an outline of important open questions that could be addressed in the future.
\end{abstract}

%\keywords{...}
%\submitto{\RPP}
\maketitle

\section{Introduction}

In our everyday lives, we are familiar with matter existing in three different states: solid, liquid, or gas. However, under more extreme conditions other states of matter can emerge. Examples of such states include plasmas at high fields and temperatures, and superfluids at extremely low temperatures. These different states of matter are defined by their distinct properties emerging from the interactions between their constituent particles. 

Liquids, for example, exhibit an ability to flow, are \textnew{nearly} incompressible, have surface tension, and are able to form droplets \cite{LandauLifshitz}. In ordinary liquids, these properties arise from a precise interplay of an intrinsic inter-particle attraction and a short-range repulsion that emerges at higher densities. While the former typically has its origin in van der Waals interactions or hydrogen bonding between the particles forming the liquid, the latter is a result of the Pauli exclusion principle that becomes important as the density of a liquid is increased and wave functions of individual particles start to overlap \cite{Pauli1940}.

While classical liquids turn into solids at low temperatures, this is not necessarily the case for quantum liquids \cite{LeggettBook}. A prominent example of this is helium, which remains a liquid near absolute zero temperature and under atmospheric pressures because of the large zero-point motion of the constituent atoms. In this temperature range quantum mechanical effects dominate, leading to the emergence of superfluidity \cite{Kapitsa1938, Allen1938}. Similar to a classical liquid, helium can also form droplets \cite{Toennies2001, Dalfovo2001, Harms1998, Toennies2004, Choi2006}. These ultracold helium droplets can be used to trap and cool dopant molecules \cite{Grebenev1998, Hartmann1995}. %Using spectroscopic methods, properties of both the droplet itself and the molecules inside the droplets can be probed. With this method properties like the low temperature of the droplets \cite{Hartmann1995} or the unimpeded rotation of dopant molecules \cite{Grebenev1998} have been experimentally established. 
With the experimental realization of Bose-Einstein condensation (BEC) \cite{Anderson1995, Davis1995, Bradley1995} and Fermi degeneracy \cite{DeMarco1999} in ultracold atomic vapours, new experimental platforms have emerged that allow us to study quantum phenomena and exotic states of matter with precise controllability and in completely novel parameter regimes \cite{Bloch2008}. %These systems can only be realized by working with ultra-dilute samples, which have to be cooled to temperatures in the range of a few tens of nanokelvin to reach the quantum degenerate regime, well below the freezing points of typical liquids \cite{Ketterle1999}.

In this review article we discuss how two new states of matter can emerge from such ultracold atomic gases as a result of fine-tuned inter-particle interactions. We elaborate on how a novel type of liquid droplets can form if two independent interactions are present in an ultracold atomic system. These quantum droplets exhibit exceptionally low densities that are eight orders of magnitude lower than that of liquid helium droplets and they exist at temperatures that are about nine orders of magnitude lower than the freezing points of classical liquids. In stark contrast to ordinary liquids, this quantum liquid droplet state of matter is formed by the interplay of weak mean-field attraction and repulsive quantum fluctuations \cite{Petrov2015}. 

Specifically, we review and compare two different systems in which the condition of two independent interactions can be met. Systems featuring both contact and magnetic dipole-dipole interactions can balance a combined attractive mean-field with repulsive quantum fluctuations \cite{Kadau2016, Ferrier-Barbut2016, Ferrier-Barbut2016a, Schmitt2016, Chomaz2016, Ferrier-Barbut2018b, Bottcher2019a}. Alternatively, homo- and heteronuclear atomic mixtures feature both inter- and intra-species interactions so droplets are formed by balancing quantum fluctuations and attractive inter-species interactions \cite{Cabrera2018, Semeghini2018, Cheiney2018, Errico2019, Ferioli2019, Ferioli2019a, Rakshit2019}. Despite sharing a similar droplet formation mechanism, the emerging quantum states in these two systems show distinct properties according to the interactions at play.

We then discuss how dipolar quantum droplets can self-assemble into coherent arrays. The recent observation of such arrays constitutes the first conclusive evidence for a supersolid state of matter arising from purely intrinsic inter-particle interactions \cite{Tanzi2019, Bottcher2019, Chomaz2019, Guo2019, Tanzi2019a, Natale2019, Hertkorn2019, Tanzi2019b, Ilzhofer2019a, Petter2020}. A supersolid is a counterintuitive superposition state that features both the crystalline structure of a solid and the frictionless flow of a superfluid \cite{Boninsegni2012, Chan2013, Leggett1970, Boninsegni2012a, Balibar2008, Prokofev2007, Galli2008, Balibar2010, Josserand2007, Kuklov2011, Saccani2012, Yukalov2020, Kora2019}. In this state, every constituent atom is part of the solid and the superfluid simultaneously. While the notion of supersolidity was introduced more than 60 years ago \cite{Penrose1956, Gross1957, Thouless1969, Andreev1969, Leggett1970, Boninsegni2012}, the direct observation of such a state has so far been limited to systems where the structure formation was mediated by external light fields \cite{Li2017, Leonard2017, Leonard2017a, Morales2018}. 

%Both of the liquid quantum droplets and the dipolar supersolids arise from a precise interplay of two independent interactions and are stabilized by purely quantum mechanical effects. 
%The interplay of two independent interactions and quantum fluctuations in both the liquid quantum droplets and the dipolar supersolids leads to many particular properties, which we will discuss in the following. In particular, we will briefly summarize the corresponding theoretical background of these states and then focus on the experimental efforts aiming to understand these new states of matter with fine-tuned interactions. We will conclude by discussing some of the many exciting open questions, prospects and experimental challenges of this emerging research area.

The interplay of two independent interactions and quantum fluctuations in liquid quantum droplets and in dipolar supersolids leads to unique properties. In the following, we discuss the emergence of these properties in each system. We briefly summarize the theoretical background and then focus on the experimental efforts aiming to understand these new states of matter with fine-tuned interactions. We conclude by discussing some of the exciting open questions, prospects and experimental challenges of this emerging research area.

\section{Theoretical description}

\subsection{Mean-field description of a BEC}

Bose-Einstein condensation requires no inter-particle interactions because the condensation process is purely a quantum statistical effect. However, the presence of interactions between the individual atoms strongly influences the properties of a BEC. In the ultracold and dilute regime, the inter-particle contact interaction potential during a scattering event is well approximated by a zero-range interaction in the form $V(r) = g \, \delta(r)$. Here, $r$ is the inter-atomic separation distance, $\delta(r)$ the Dirac delta function, and $g = 4\pi \hbar^2 \as / m$ the coupling constant of the contact interaction with the scattering length $\as$ and the atomic mass $m$. 

Within a mean-field approximation, the energy per particle for a uniform BEC with a density $n$ is then given by \textold{$E/N = 1/2 \, g n$}\textnew{$E/N = g n/2$} \cite{PitaevskiiStringariBook}. This simple equation already reveals two important properties: for attractive interaction ($g < 0 $) the system collapses on itself, and for repulsive interactions ($g > 0$) a BEC can only exist in the gaseous phase where the energy per particle is minimized at the lowest density. Therefore, experiments with BECs need to be performed in an external trapping potential $V_\mathrm{ext}(r)$ to localize the atomic cloud and to keep it from expanding. \textnew{The presence of an external trap can furthermore also stabilize a condensate against collapse for weak attractive interactions \cite{Dalfovo1999}.}

Going beyond the mean-field approximation leads to corrections to the ground state energy stemming from quantum fluctuations of the collective modes in a BEC. The leading-order correction term was first calculated in 1957 by Lee, Huang, and Yang and is thus called the LHY-correction \cite{Lee1957}. In this framework, the ground state is composed of a large fraction of atoms in the condensate and a small fraction of atoms in excited states. This beyond mean-field correction leads to a small shift in the energy per particle that increases with the interaction strength $g$ and exhibits a stronger density dependence than the mean-field energy. The corrected energy per particle of the ground state is then given by \textold{$E/N = 1/2 \, g n + \alpha_\mathrm{LHY}\, g\,(gn)^{3/2}$}\textnew{$E/N = g n/2 + \alpha_\mathrm{LHY}\, g\,(gn)^{3/2}$} \textnew{, where $\alpha_\mathrm{LHY}$ is the numerical prefactor of the beyond mean-field correction}. 

\subsection{Quantum droplets}

A way to realize a novel and liquid-like state in such a system was proposed by D. Petrov in 2015 \cite{Petrov2015} for Bose-Bose mixtures, and experimentally discovered independently shortly afterwards by our group in Stuttgart in an ensemble of ultracold dipolar atoms \cite{Kadau2016, Ferrier-Barbut2016}. In 2018 quantum droplets in a Bose-Bose mixture were observed for the first time \cite{Cabrera2018}. The experimental platforms considered in these studies -- Bose-Bose mixtures and dipolar quantum gases -- initially appear to be distinct platforms. However, both realize a bosonic system with two independent interactions. By fine-tuning the interplay between these two interactions, new states of matter can be realized.

\begin{figure}[t!]
\begin{center}
\begin{overpic}[width=0.95\textwidth]{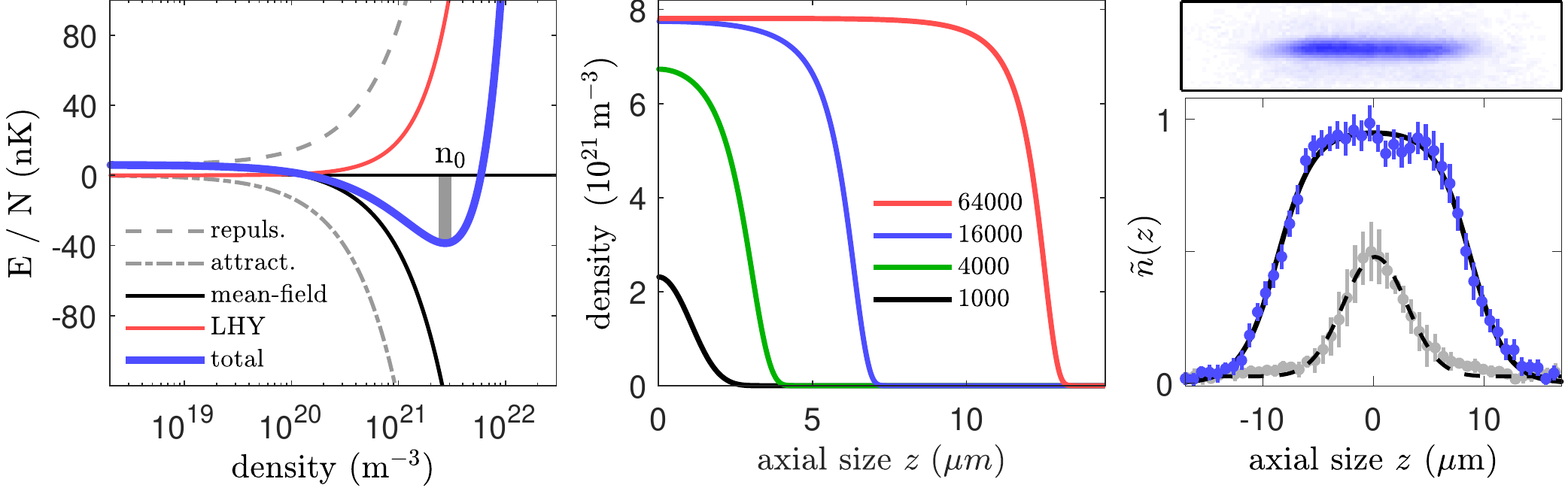}
\put(1.5, 30){\textbf{a}} \put(37, 30){\textbf{b}} \put(72.5, 30){\textbf{c}}
\end{overpic}
\caption{
(\textbf{a}) Illustration of the different contributions to the total energy per particle. \textnew{The contributions of the attractive (dash-dotted) and repulsive (dashed) interaction result in an overall weakly attractive mean-field energy (black). This}\textold{The weak} attractive mean-field energy \textold{(black) }is balanced by repulsive quantum fluctuations (red) at a finite equilibrium density $n_0$, leading to a minimum in the total energy per particle (blue).
(\textbf{b}) Characteristic axial density profile of a self-bound dipolar quantum droplet \textnew{consisting of $^{164}$Dy atoms} for different atom numbers\textnew{ and a scattering length of $a_\mathrm{s} = 70\, a_0$}. For high atom numbers the peak density of the droplet saturates. Increasing the atom number further only leads to an increase in the size, but not in the peak density of the droplet. \textnew{It is important to note, that the density saturation in dipolar quantum droplets only occurs along the axial direction due to the anisotropic nature of the dipole-dipole interaction.}
%Characteristic density profile $|\phi_0(r)|^2$ of a self-bound quantum droplet in a Bose-Bose mixture for different atom numbers defined by $\tilde{N}$. For high atom numbers the peak density of the droplet saturates and further increasing the atom number only leads to an increase in the size, but not in the peak density of the droplet. Adapted from \cite{Petrov2015}.
(\textbf{c}) Mean density profile and normalized cut through the center of this density profile for a dipolar quantum droplet of $^{162}$Dy with $N \approx 2.2 \times 10^4$ and $\as = 92(4)\,a_0$. Shortly after the formation of the droplet, a flat-top density distribution (blue) can be observed. Due to rapid three-body loss the droplet is in a highly dynamical state and for longer evolution times (gray) the density saturation rapidly vanishes. Figures adapted from \cite{Bottcher2020}.
}
\label{fig:Schematic}
\end{center}
\end{figure}

Quantum droplets can arise for competing interactions, where one of the interactions is attractive \textnew{($g_\mathrm{att}$)} and the other repulsive \textnew{($g_\mathrm{rep}$)}. In this case, the mean-field energy depends on the difference of the two coupling constants \textold{$\delta g = |g_1| - |g_2|$}\textnew{$\delta g = |g_\mathrm{rep}| - |g_\mathrm{att}|$} of the two interactions and can be significantly reduced. In contrast to this, the beyond mean-field correction stemming from quantum fluctuations depends on the individual coupling constants and can thus be effectively enhanced by the presence of the two interactions. In the case of a weakly attractive combination of interactions, a repulsive beyond mean-field correction can therefore stabilize the system. This stabilization is illustrated in \Fref{fig:Schematic}\textbf{a}, where the droplet density corresponds to the minimum of the energy per particle. At this energy minimum the attractive mean-field and the repulsive beyond mean-field correction balance each other. Around this equilibrium density, a stable state can be formed. This new state corresponds to liquid quantum droplets. 

\Fref{fig:Schematic}\textbf{b} illustrates the saturation of the droplet peak density with increasing atom number \cite{Petrov2015, Bottcher2020}. For large enough atom numbers we observe the liquid-like saturation of the peak density and increasing the number of particles only leads to an increase in the size of the droplet.  The density profiles in \Fref{fig:Schematic}\textbf{b} were calculated using the current state-of-the-art description of quantum droplets, based on an extended Gross-Pitaevskii equation (eGPE). 

The eGPE includes the well-known kinetic energy, external trapping, and two-body interactions described by the Gross-Pitaevskii equation \cite{PitaevskiiStringariBook}. Furthermore, the eGPE includes the beyond mean-field correction stemming from the repulsive quantum fluctuations. It is important to note that this beyond mean-field correction has only been calculated for a homogeneous system and can therefore only be included within a local-density approximation. Studies in which the full many-body system has been solved using quantum Monte-Carlo (QMC) methods have verified the formation of quantum droplet states and are in reasonable agreement with the predictions of the eGPE framework, both for Bose-Bose mixtures and dipolar quantum gases \cite{Saito2016, Macia2016, Cinti2017, Cinti2017a, Bombin2017, Cikojevic2018, Cikojevic2019, Parisi2019}. 
%Quantum droplets have been experimentally realized in Bose-Bose mixtures with repulsive inter-species and attractive intra-species interactions \cite{Cabrera2018, Semeghini2018, Cheiney2018, Errico2019, Ferioli2019, Ferioli2019a} and in quantum gases of atomic species with a sizeable magnetic dipole moment $\mu_\mathrm{m}$ \cite{Kadau2016, Ferrier-Barbut2016, Ferrier-Barbut2016a, Schmitt2016, Chomaz2016}.
In the following sections we briefly discuss the eGPE theory for both of these systems.

\subsection*{Quantum droplets in Bose-Bose mixtures}

As we have seen above, the attractive mean-field and the repulsive beyond mean-field contribution balance each other for the equilibrium density of a quantum droplet. In general, the different intra-species scattering lengths $a_{11}$ and $a_{22}$ lead to different equilibrium densities $n_0^{(i)}$ for the two components of the mixture. In this situation, the droplet state forms with an intrinsic imbalance in the atom numbers of the two components that is given by $N_1/N_2 = \sqrt{a_{22} / a_{11}}$. Furthermore, under the assumption of the same spatial mode $\Psi_1 = \sqrt{n_1} \, \psi$ and $\Psi_2 = \sqrt{n_2}\, \psi$ for the two components of a homonuclear mixture \cite{Petrov2015, Cheiney2018}, the eGPE can be brought into the form

\begin{equation}\label{eq:BBeGPE}
i\hbar \,\partial_t \psi = \bigg[ -\frac{\hbar^2 \nabla^2}{2m} + V_\mathrm{ext}(\vec{r}) + \alpha\, n_0 \, |\psi|^2 + \gamma  \, n_0^{3/2} \, \left| \psi \right|^3 \bigg] \,\psi \;.
\end{equation}
Note that the wavefunction $\psi = \psi (\vec{r},t)$ in this equation depends on the spatial coordinates $\vec{r}$ and time $t$. Here, $n_0 = n_1 + n_2$ is the total peak density of the mixture. The mean-field contribution in the eGPE is defined by
\begin{equation}\label{eq:BBalpha}
\alpha = \frac{8 \pi \hbar^2}{m} \, \frac{\sqrt{a_{11}/a_{22}}}{\left(1 + \sqrt{a_{11}/a_{22}}\right)^2} \; \delta a\; ,
\end{equation}
with the repulsive intra-species scattering lengths $a_{11}$ and $a_{22}$, the attractive inter-species scattering length $a_{12}$ and the effective scattering length $\delta a = a_{12} + \sqrt{a_{11} a_{22}}$. The pre-factor of the quantum fluctuation contribution for Bose-Bose mixtures is given by
\begin{equation}\label{eq:BBgamma}
\gamma  = \frac{128 \sqrt{\pi} \hbar^2}{3 m} \left( \frac{\sqrt{a_{11} a_{22}}}{1 + \sqrt{a_{11}/a_{22}}}\right)^{5/2} \; f\left(\frac{a_{12}^2}{a_{11} a_{22}} , \sqrt{\frac{a_{11}}{a_{22}}}\right)\; ,
\end{equation}
with the function $f(x,y) = \sum_\pm \left(1 + y \pm \sqrt{(1-y)^2 + 4xy} \right)^{5/2} \, / 4\sqrt{2}$ \cite{Petrov2015, Larsen1963}. \textnew{Note that the quantum fluctuation term for $\delta a < 0$ features a small imaginary part, which is neglected within this theoretical framework.} To quantify the atom number in the droplet state, we furthermore introduce the parameter $\tilde{N}$ such that the atom number of the i-th component is given by $N_\mathrm{i} = n_0^{(i)} \xi^3 \tilde{N}$, with $\xi$ being a natural length scale \cite{Petrov2015}.

The resulting density of the quantum droplets is significantly higher than the density in the original BEC. This larger density increases the rate of three-body processes, leading to a faster loss of atoms from the droplets \cite{Fedichev1996a, Weber2003}. In order to understand the observed dynamics in the experiments, three-body losses have to be included in our description \cite{Wachtler2016}. This can be phenomenologically done by adding the term $(-i \hbar/2)\, L_3 \, |\psi|^4 \psi$ to the right side of \Eref{eq:BBeGPE}. \textnew{In general, the three-body loss rates for the four different combinations of atoms from the mixture can be different.}

\subsection*{Dipolar quantum droplets}

A different approach to realize quantum droplets uses quantum gases composed of an atomic species with a sizeable magnetic dipole moment $\mu_\mathrm{m}$ \cite{Lahaye2009}. In this case, the individual atoms interact through the contact interaction, as well as the dipole-dipole interaction. The dipole-dipole interaction between two polarized dipoles at the positions $\vec{r}$ and $\vec{r}^\prime$ is given by

\begin{equation}\label{eq:ddi}
V_\mathrm{dd}(\vec{r}-\vec{r}^\prime) \; = \; \frac{\mu_0 \, \mu_\mathrm{m}^2}{4\pi} \; \frac{1 - 3 \cos^2 \vartheta}{|\vec{r}-\vec{r}^\prime|^3} \;,
\end{equation}
with the angle $\vartheta$ between the polarization direction $\vec{\hat{z}}$ and the relative position of the two dipoles $|\vec{r}-\vec{r}^\prime|$. Using a mean-field approximation, the eGPE for the dipolar system is then given by

\begin{equation}\label{eq:eGPE}
i\hbar \,\partial_t \psi = \bigg[ -\frac{\hbar^2 \nabla^2}{2m} + V_\mathrm{ext}(\vec{r}) + g \, |\psi|^2 + \Phi_\mathrm{dd} + g_\mathrm{qf} \, \left| \psi \right|^3 \bigg] \,\psi \;.
\end{equation}
In this equation $g |\psi|^2$ is \textnew{related to} the mean-field contact interaction and $\Phi_\mathrm{dd}(\vec{r},t)$ is \textnew{related to} the mean-field dipolar interaction, which is defined by
\begin{equation}\label{eq:Phidd}
\Phi_\mathrm{dd}(\vec{r},t) = \int\!\mathrm{d}\vec{r}^\prime \, V_\mathrm{dd}(\vec{r}-\vec{r}^\prime) |\psi(\vec{r}^\prime,t)|^2 \;.
\end{equation}

The last term in \Eref{eq:eGPE} is the beyond mean-field correction stemming from quantum fluctuations \cite{Schutzhold2006, Lima2011, Lima2012, Cherny2019}. In the dipolar case, this correction is given by 

\begin{equation}\label{eq:gqf}
g_\mathrm{qf} = \frac{32}{3\sqrt{\pi}} \,g \, \sqrt{\as^3} \; Q_5(\edd) \; \approx \; \frac{32}{3\sqrt{\pi}}  g \, \sqrt{\as^3} \, \left( 1 + \frac{3}{2} \, \edd^2 \right) \;.
\end{equation}
Here, ${Q_5(\alpha) = 1/2 \int_0^\pi \mathrm{d}\theta \, \sin(\theta) \left[1 + \edd (3 \cos^2 \theta - 1)\right]^{5/2}}$ describes the averaged angular contribution of the dipolar interaction \cite{Lima2012}. The quantum fluctuations depend on the ratio of the interaction length scales $\edd = \add/\as$, where the length scale of the dipolar interaction is $\add = m \,\mu_0 \mu_\mathrm{m}^2/(12 \pi \hbar^2)$ and the s-wave scattering length is $\as$. Due to the emergence of a phonon instability in the homogeneous system for large dipolar strengths \cite{Lahaye2009}, this correction due to the quantum fluctuations is only valid for $\edd < 1$. For larger dipolar strength with $\edd > 1$, $g_\mathrm{qf}$ acquires a small imaginary part that is typically neglected in the theoretical description of dipolar quantum droplets.

Notably, the anisotropy of the dipolar interaction is responsible for an anisotropic binding mechanism for the quantum droplets. As a consequence, dipolar quantum droplets are elongated along the polarizing magnetic field direction. This also means that the density saturation demonstrated in \Fref{fig:Schematic}\textbf{b} only appears along the axial direction of the dipolar droplets.

Similar to Bose-Bose mixtures, three-body loss needs to be included to understand the dynamical behaviour of the dipolar quantum droplets in the experiments. \textnew{However, in the dipolar system only one three-body loss rate exists, which constitutes an important difference compared to the binary mixtures.}

\section{Experimental realization of quantum droplets}

Quantum droplets were first observed using dipolar quantum gases made of $^{164}$Dy atoms \cite{Kadau2016}. Since then quantum droplets have been realized with dipolar systems made of $^{164}$Dy \cite{Kadau2016, Ferrier-Barbut2016, Ferrier-Barbut2016a, Schmitt2016, Ferrier-Barbut2018b}, $^{162}$Dy \cite{Bottcher2019a} and $^{166}$Er \cite{Chomaz2016} atoms. Bose-Bose droplets were first observed in mixtures of different spin states of $^{39}$K \cite{Cabrera2018, Semeghini2018, Cheiney2018, Ferioli2019, Ferioli2019a}, and subsequently also in different heteronuclear mixtures of $^{41}$K-$^{87}$Rb \cite{Errico2019, Burchianti2020} and $^{23}$Na-$^{87}$Rb \cite{Guo2020}.

In the following, we discuss the properties of these different quantum droplets and highlight some similarities to other liquids.

\subsection{Liquid-like density saturation}

In \Fref{fig:Schematic}\textbf{b} we have seen the theoretical prediction of a liquid-like saturation of the peak density of a quantum droplet. Such a saturation is a characteristic property of \textold{an incompressible}\textnew{a} liquid \textnew{with a very low compressibility} that wants to maintain a constant density.

Indications of this density saturation have been observed in experiments with trapped quantum droplets made of $^{162}$Dy atoms \cite{Bottcher2020}. By preparing a stable BEC and subsequently ramping the scattering length into the droplet regime, these experiments showed a single elongated quantum droplet containing approximately $2.2\,\times\,10^4$ atoms. \Fref{fig:Schematic}\textbf{c} illustrates the density profile of such a droplet. A cut through the center of this profile reveals the characteristic flat-top density distribution, which agrees well with the density distribution known from liquid helium droplets \cite{Harms1998}. 

The observed flattening of the axial density profile of a dipolar quantum droplet only persists for a short time because the corresponding density is high, which consequently leads to rapid three-body losses of atoms from the droplet. This effect can be seen for droplets that have undergone an additional evolution time of \unit[5]{ms} (gray points in \Fref{fig:Schematic}\textbf{c}). At this point about half of the initial atoms remain and the droplet can be well described by a Gaussian density distribution. With the significantly reduced axial size, no trace of the previously observed density saturation remains. We emphasize that the observed density profile is not the density profile \textnew{of} the quantum droplet ground state, but the rapid losses make it a highly dynamical state. Theoretical studies have shown a pathway to create larger and less excited droplet states, which can be used in future experiments to further study the liquid-like properties of the droplets \cite{Mennemann2019}.

\subsection{Critical atom number of a self-bound quantum droplet}

Arguably the most liquid-like aspect that arises from the balance between attractive mean-field interaction and repulsive quantum fluctuations is the self-bound nature of the quantum droplets \cite{Petrov2015, Baillie2016, Wachtler2016a}. A water droplet is an example of a self-bound system that does not need a container to remain bound. This is in stark contrast to gases which expand freely if they are not confined by an external container. 

Ultracold atomic systems like BECs and degenerate Fermi gases constitute gaseous states that expand upon releasing them from the external trapping potential. Quantum droplets, on the other hand, are self-bound if a critical atom number is exceeded. This critical atom number arises from the interplay between the binding mechanism and the kinetic energy cost of an inhomogeneous density distribution. The associated kinetic energy leads to an increase in the energy per particle, which for small atom numbers can be strong enough to drive a liquid-to-gas transition.

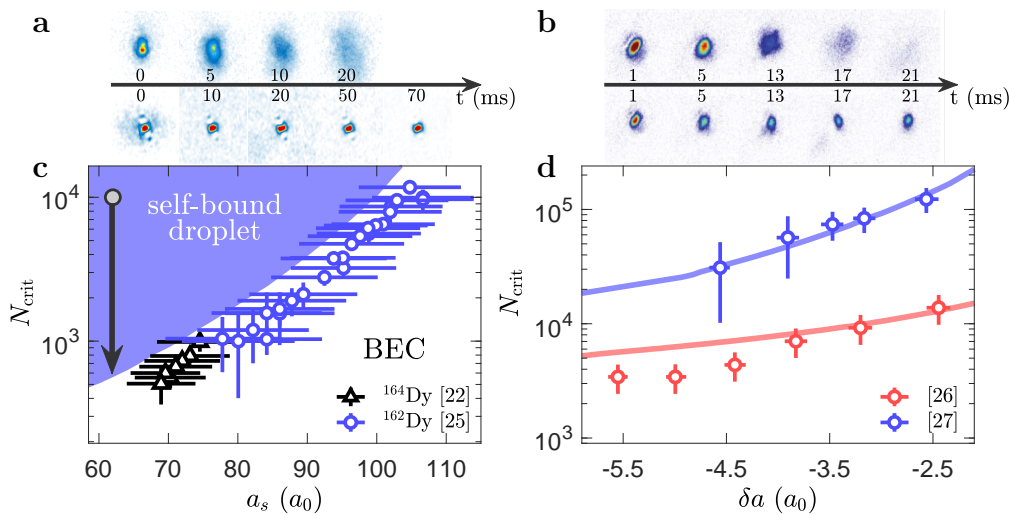
\begin{figure}[t!]
\begin{center}
\begin{overpic}[width=0.95\textwidth]{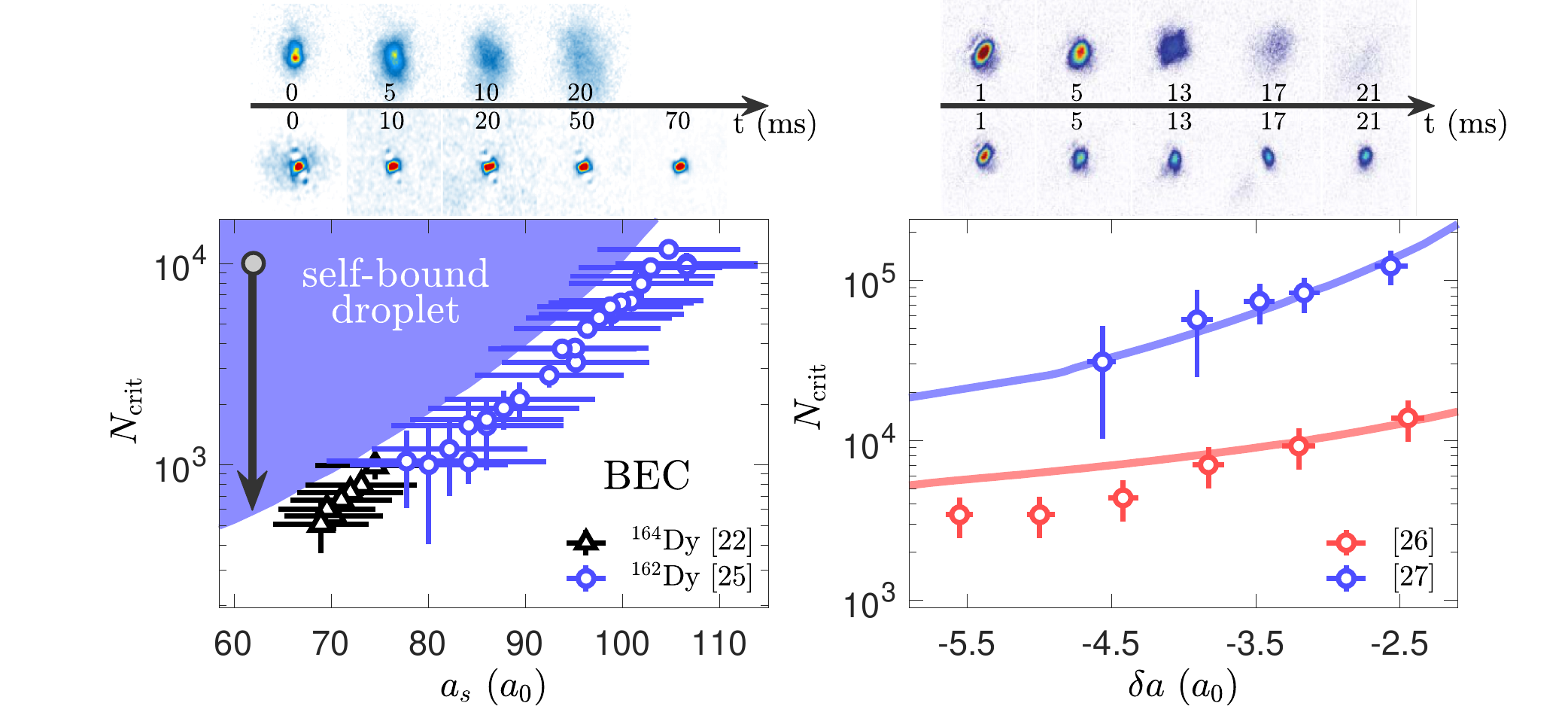}
\put(9, 43){\textbf{a}} 
\put(54, 43){\textbf{b}} 
\put(9, 30){\textbf{c}} 
\put(54, 30){\textbf{d}}
\end{overpic}
\caption{
Example images for different expansion times of a BEC and a self-bound quantum droplet in free space for (\textbf{a}) the dipolar and (\textbf{b}) the Bose-Bose mixture. \textnew{In the droplet regime, imaging aberations appear due to the high density and a spatial size of the droplets that is smaller than the experimental imaging resolution.}
Theoretical phase boundary between self-bound droplet and expanding BEC, together with the measured critical atom numbers for (\textbf{c}) the dipolar \cite{Schmitt2016, Bottcher2019a} and (\textbf{d}) Bose-Bose droplets \cite{Cabrera2018, Semeghini2018}. In \cite{Cabrera2018} a residual confinement along the vertical direction was used, leading to the lower critical atom number compared to \cite{Semeghini2018}. In \textbf{c} the measurement strategy to determine the critical atom number is indicated by the black arrow (see main text). Images and data adapted from \cite{Schmitt2016, Bottcher2019a, Bottcher2020} and \cite{Cabrera2018, Semeghini2018}.
}
\label{fig:CritNumber}
\end{center}
\end{figure}

The self-bound nature of these quantum droplets has been experimentally demonstrated in dipolar systems with $^{164}$Dy \cite{Schmitt2016} and $^{162}$Dy \cite{Bottcher2019a}, in homonuclear Bose-Bose mixtures of different spin states in $^{39}$K \cite{Cabrera2018, Semeghini2018}, and in heteronuclear mixtures of \textold{ $^{41}$K and $^{87}$Rb}\textnew{$^{41}$K-$^{87}$Rb \cite{Errico2019, Burchianti2020} and $^{23}$Na-$^{87}$Rb \cite{Guo2020}}. Example images of self-bound droplets and expanding BECs for different expansion times are shown in \Fref{fig:CritNumber}\textbf{a} and \textbf{b} for the dipolar system and the Bose mixture, respectively. Closely related behaviour has also been observed in dipolar $^{166}$Er, where the expansion velocity of the system was significantly reduced in the droplet regime \cite{Chomaz2016}. However, in this case the three-body loss limited lifetime of the droplets was too short for a direct observation of the self-bound nature. 

Different measurements of the critical atom number all use a similar experimental sequence, which is illustrated by the black arrow in \Fref{fig:CritNumber}\textbf{c}. After preparing a quantum droplet with a high atom number, the external trapping potential is switched off and the unavoidable process of three-body decay leads to a rapid loss of atoms. Upon crossing the phase boundary to the gaseous state, the droplet turns into a gas and rapidly expands. This expansion leads to a significant reduction in the density, thus suppressing further losses. As a result, atom number decay curves are observed that settle at a constant atom number -- the critical atom number of a self-bound droplet. 
 
The actual value of the critical atom number of a self-bound droplet depends on the precise strengths of the two interactions involved. By varying the effective scattering length using Feshbach resonances, the phase diagrams shown in \Fref{fig:CritNumber}\textbf{c} and \textbf{d} can be mapped out for the dipolar system and the Bose-Bose mixtures, respectively. The measurements with Bose-Bose droplets presented in \cite{Cabrera2018} use a residual confinement along the direction of gravity which leads to a significant reduction of the critical atom number. The phase boundary is qualitatively reproduced using the eGPE framework. However, in both systems discrepancies between experiment and theory are visible, which is discussed in the last chapter of this review.

\subsection{Collective Excitations}

Collective excitations are a crucial concept in many areas of physics and their measurement can act as a sensitive probe for the underlying phenomena. In the case of quantum droplets, the different interactions at play lead to specific excitations in each of the two experimental systems.

A peculiar feature of quantum droplets in Bose-Bose mixtures are the two distinct components, which allows for an excitation in which they move either in- or out-of-phase with respect to each other. \textold{However, so far no collective modes have been observed in Bose-Bose droplets due to their short experimental lifetime.}
Notably, collective modes would lead to exotic finite temperature behaviour of the \textnew{binary} droplets. In his seminal work on droplets \cite{Petrov2015}, D. Petrov showed that \textnew{in a three-dimensional system} there exists a regime where no collective modes are below the chemical potential, which corresponds to the particle emission threshold. This means that exciting the quantum droplet would lead to the spilling of particles, and hence to a self-evaporation of the droplet to zero temperature. \textnew{In particular the out-of-phase modes are expected to be overdamped because they are too energetic to be bound. So far, no collective modes have been observed in Bose-Bose droplets due to their short experimental lifetime.} 

This self-evaporation is a specific feature of quantum droplets in Bose-Bose mixtures, since in the dipolar case at least one collective mode is always expected to remain below the particle emission threshold \cite{Baillie2017}. However, an observation of the self-evaporation would require the development of novel temperature probes. This is due to the fact that the quantum droplets are self-bound objects, meaning that the usual time-of-flight expansion temperature measurements for cold atoms cannot be applied \cite{Ketterle1999, Wenzel2018a}. %One potential example of such a new probe, which is inspired by the rotational spectroscopy of molecules in helium droplets \cite{Hartmann1995, Grebenev1998}, is the trapping and probing of impurities inside a quantum droplet \cite{Wenzel2018a}.
In contrast to studies on three-dimensional systems, a theoretical study of the collective modes in a one-dimensional quantum droplet showed that the breathing mode always stays trapped \cite{Tylutki2020}.

\begin{figure}[t!]
\begin{center}
\begin{overpic}[width=0.93\textwidth]{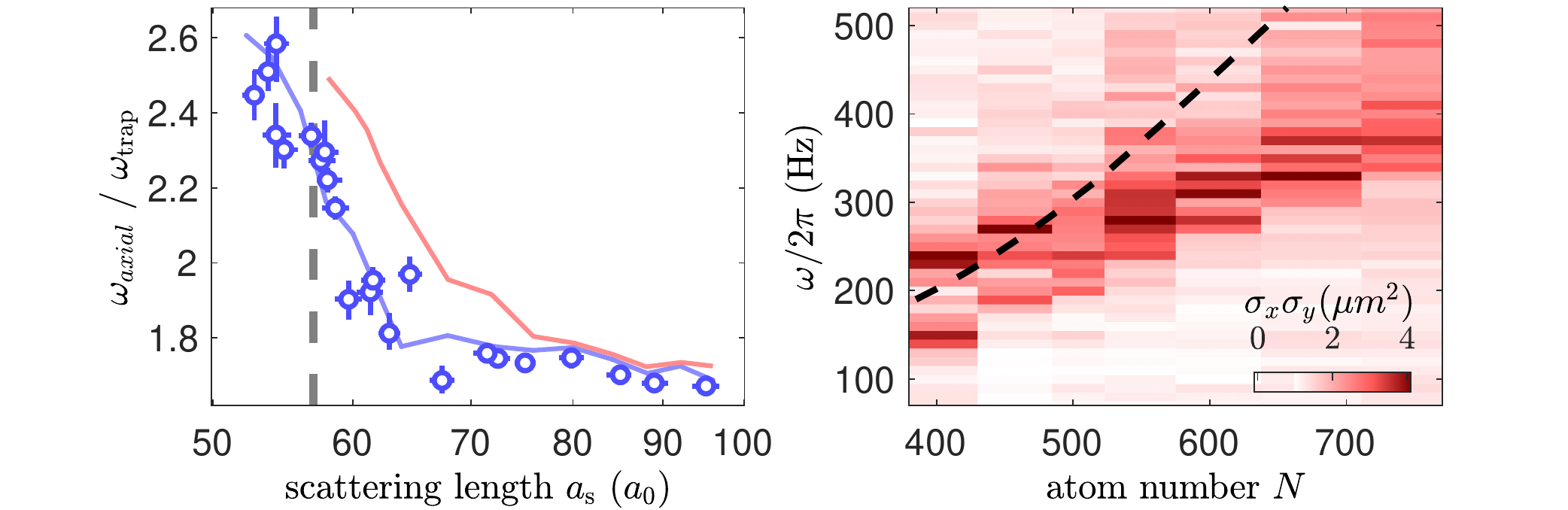}
\put(6, 31.5){\textbf{a}} \put(50, 31.5){\textbf{b}}
\end{overpic}
\caption{Measurements of the collective excitations of a dipolar quantum droplet. (\textbf{a}) The excitation frequency of the axial quadrupole mode $\omega_\mathrm{axial}$ in the crossover from regular BEC to quantum droplet \cite{Chomaz2016}. The results can be well reproduced with numerical simulations that include the stabilizing contribution of quantum fluctuations (blue line), while simulations without this stabilization (red line) show a significant discrepancy.
(\textbf{b}) Measurement of the scissors mode arising from the breaking of the rotational symmetry in dipolar quantum droplets. \textnew{Experimentally, the orientation of the magnetic field is modulated at different frequencies and the response of the system is observed by looking at the droplet sizes $\sigma_\mathrm{x,y}$. The dashed line shows the theoretical scissors mode frequency expected from linear response theory.}
Data taken from \cite{Chomaz2016} and \cite{Ferrier-Barbut2018b}, respectively.
}
\label{fig:CollectiveExcitations}
\end{center}
\end{figure}

In dipolar systems, measurements of collective excitations have been performed shortly after the discovery of dipolar quantum droplets \cite{Baillie2017, Odziejewski2016}. Experiments with erbium focused on the lowest-lying quadrupole mode in the crossover between regular BEC and dipolar quantum droplet \cite{Chomaz2016}. This excitation mainly leads to an oscillation in the axial length of the droplet\textold{, as illustrated in the inset of \Fref{fig:CollectiveExcitations}\textbf{a}}. The measurements shown in \Fref{fig:CollectiveExcitations}\textbf{a} clearly show the increase in the excitation energy upon approaching the transition, which is well captured by numerical simulations that include quantum fluctuations as the stabilization mechanism for the droplets.

As we have seen above, the anisotropic nature of the dipolar interaction leads to elongated droplets. This elongation facilitates an angular oscillation of the droplet around the polarizing magnetic field direction, the scissors mode \cite{Ferrier-Barbut2018b}. The scissors mode is \textnew{normally} an important marker of superfluidity because superfluids are irrotational and their moment of inertia differs from the classical rigid-body value \cite{PitaevskiiStringariBook}. \textnew{However, the corresponding moment of inertia of the quantum droplets does not differ significantly from the classical rigid value due to the large anisotropy in their density distribution.} The measured response to an excitation of the angle of the polarizing magnetic field is shown in \Fref{fig:CollectiveExcitations}\textbf{b}. In these measurements, the large excitation amplitude led to a non-linear coupling of the scissors mode to other low-lying modes.

\subsection{Crossover from soliton to droplet}

In lower dimensions, Bose-Einstein condensates can also support another type of self-bound object, the so-called soliton \cite{Dauxois2006}. Solitons are well-known from various fields ranging from solitary waves in water channels to optical solitons in non-linear media. The latter ones have found important applications in telecommunication \cite{Dauxois2006}. Both bright and dark solitons have been studied extensively using Bose-Einstein condensates with attractive contact interactions \cite{Burger1999, Denschlag2000, Strecker2002, Khaykovich2002, Strecker2003, Cornish2006, Marchant2013, Medley2014, McDonald2014, Lepoutre2016, Kartashov2019}. 

The connection between \textold{such}\textnew{bright} solitons and droplets has been investigated in a work by the group of L. Tarruel, where they studied the crossover between the two using a Bose-Bose mixture in an optical waveguide \cite{Cheiney2018}. In the experiment, the attractive interactions of the mixture facilitates the realization of both \textnew{bright} solitons and quantum droplets.

While the emergence of solitons is essentially a single particle effect where the dispersion in the waveguide stabilizes the system against collapse, quantum droplets are stabilized by a many-body phenomenon, namely the repulsive quantum fluctuations. For this reason, solitons and quantum droplets are expected to exhibit fundamentally different behaviour. While solitons are only stable as long as the gas remains effectively one-dimensional, quantum droplets can exist as a purely self-bound state in three dimensions. However, the latter requires the atom number to be above the critical atom number, while the low-dimensional nature of solitons limits their existence to be below a certain maximal atom number. 

\begin{figure}[t!]
\begin{center}
\begin{overpic}[width=0.95\textwidth]{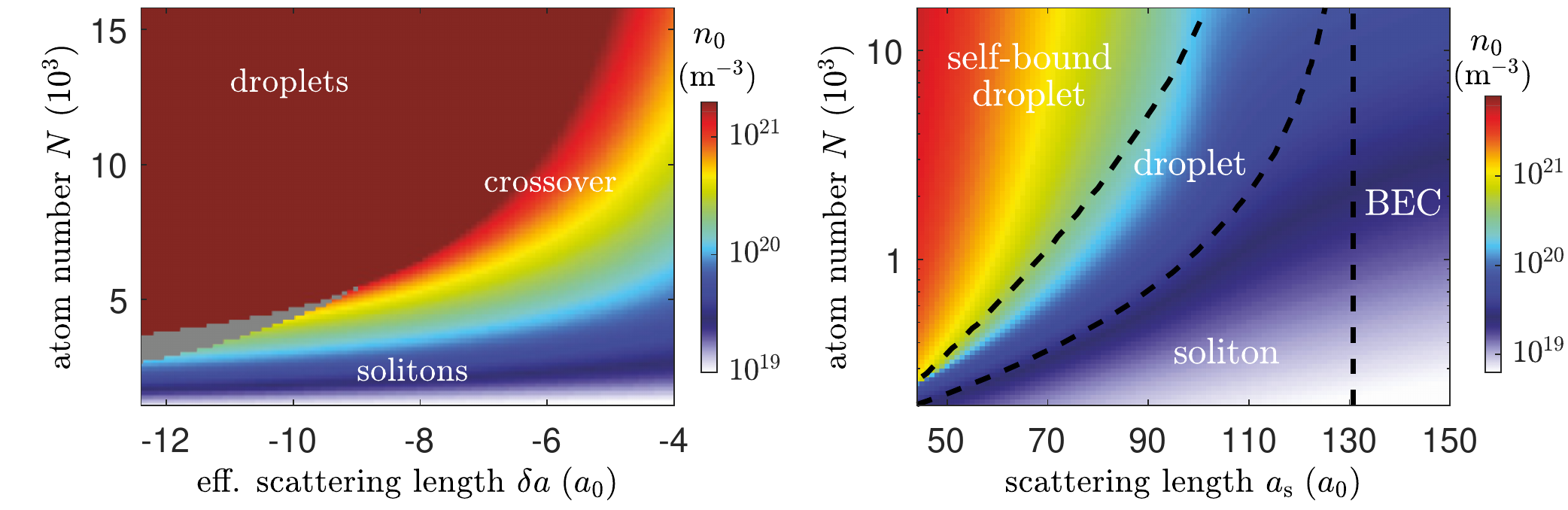}
\put(4, 31.5){\textbf{a}} \put(50, 31.5){\textbf{b}}
\end{overpic}
\caption{Simulated phase diagram of the ground-state peak density as a function of the atom number and the (effective) scattering length for a (\textbf{a}) \textold{one-dimensional }Bose-Bose mixture \textnew{in a cigar-shaped confinement} and (\textbf{b}) a \textold{two-dimensional} dipolar quantum gas\textnew{ in a pancake-shaped confinement}. Quantum droplets and solitons are distinct solutions of the eGPE, which exist for high and low atom numbers, respectively. In the intermediate range of atom numbers, the two solutions can coexist in a bistable region (gray area in (\textbf{a})) or can be smoothly connected by a crossover. \textnew{The dashled lines in (\textbf{b}) act as a guide to the eye separating the different regimes. In the dipolar case, we further differentiate between trap-bound and self-bound quantum droplets.}
Adapted from \cite{Cheiney2018} and \cite{Ferrier-Barbut2018b}, respectively.
}
\label{fig:soliton}
\end{center}
\end{figure}

These properties are key to understand the \textold{crossover}\textnew{observations} in the experiment, where the observed self-bound states evolved from soliton-like to droplet-like as a function of atom number. As it can be seen in \Fref{fig:soliton}\textbf{a}, depending on the given values of interaction strength and confinement, the two regimes were observed to be smoothly connected or remained distinct.

A similar \textold{crossover}\textnew{transition} between soliton and droplet regime is also expected to occur in dipolar quantum gases. In contrast to the purely contact-interacting system where solitons require a one-dimensional system, dipolar quantum gases can also feature solitons in two-dimensional geometries \cite{Pedri2005, Tikhonenkov2008, Koberle2012}. The \textold{crossover}\textnew{transition} from solitons at low atom numbers to quantum droplets at larger atom numbers has been theoretically studied in \cite{Ferrier-Barbut2018b} and is shown in \Fref{fig:soliton}\textbf{b}. \textnew{In their work, the authors furthermore differentiate between droplets that are only bound in the presence of an external confinement, and self-bound droplets that remain bound even in free space.}

\section{Experiments with multiple droplets}

\subsection{Arrays of dipolar quantum droplets} \label{ch:arrays}

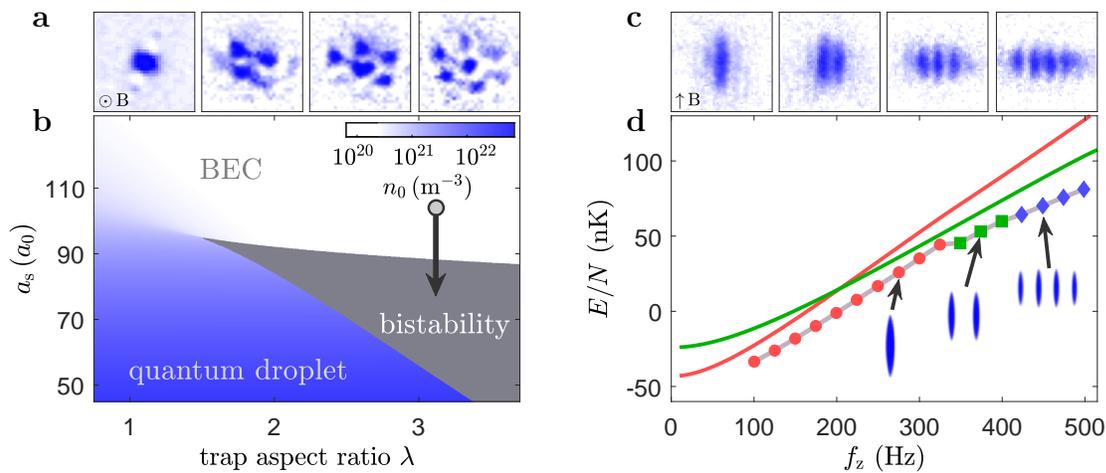
\begin{figure}[t!]
\begin{center}
\begin{overpic}[width=0.95\textwidth]{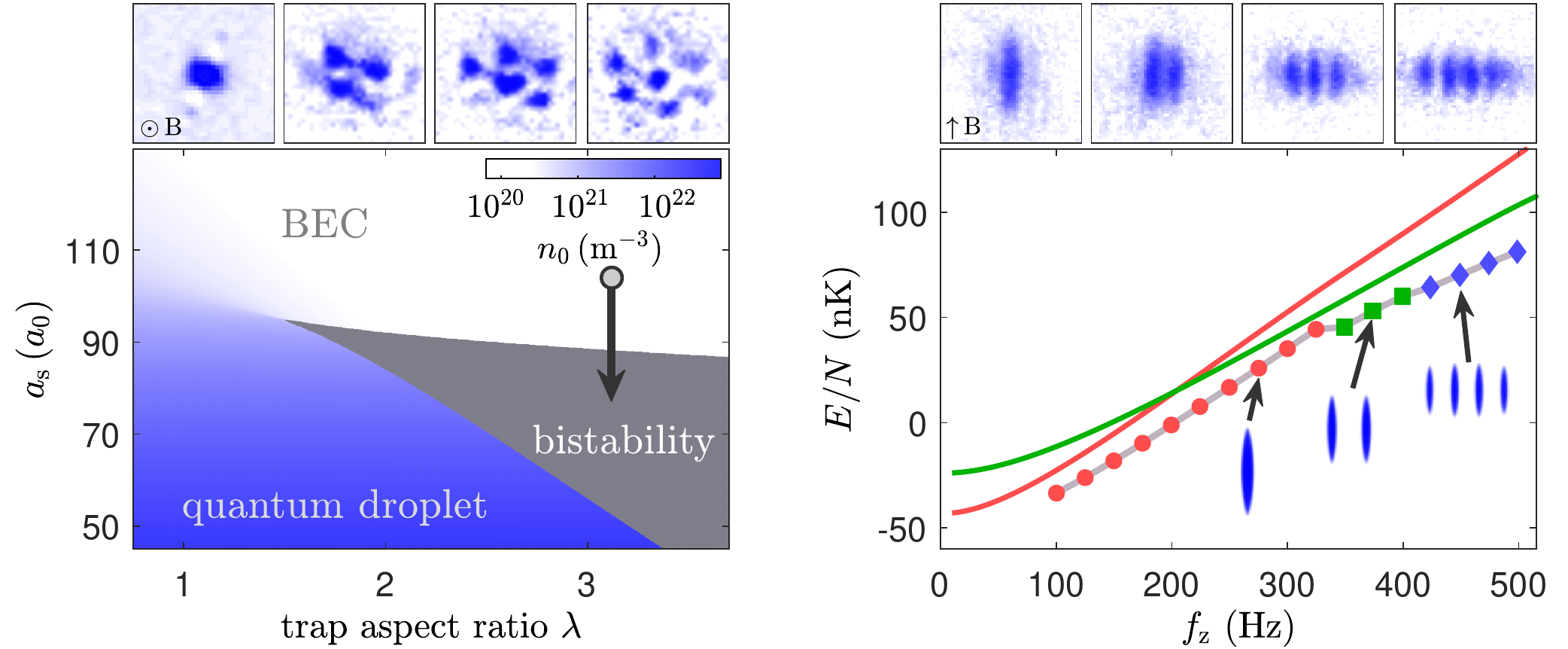}
\put(3, 39.8){\textbf{a}} \put(56, 39.8){\textbf{c}}
\put(3, 30.5){\textbf{b}} \put(56, 30.5){\textbf{d}}
\end{overpic}
\caption{
(\textbf{a}) Example images of two-dimensional droplet arrays with increasing atom numbers in a cylindrically symmetric trap \cite{Kadau2016, Schmitt2017}. 
(\textbf{b}) Phase diagram of an ensemble of $2.5\times10^4$ dysprosium atoms in a cylindrically symmetric trap with a trap aspect ratio $\lambda = \omega_\mathrm{z} / \omega_\mathrm{r}$. For the case of $\lambda < \lambda_\mathrm{c}$ there is a continuous transition from a regular BEC to a single quantum droplet. For $\lambda > \lambda_\mathrm{c}$, the BEC and droplet solutions coexist in a certain range of the scattering length, leading to the bistable region as indicated by the gray color.
(\textbf{c}), (\textbf{d}) In a cigar-shaped trap geometry, it is energetically favourable for the droplet number to increase with increasing confinement along the polarization direction. (\textbf{c}) Shows example images for this behaviour \textnew{upon changing the in-plane trap geometry} \cite{Wenzel2017}.
(\textbf{d}) Total energy per particle $E/N$ for a single droplet (red), two droplets (green) and four droplets (blue) in an external trap with $\omega_\mathrm{trap} = 2\pi \, (70, 1000, f_\mathrm{z})\,\mathrm{Hz}$ using a variational approach (lines) or numerical simulations of the eGPE (points). Data adapted from \cite{Kadau2016, Wenzel2017, Bottcher2020}.
}
\label{fig:arrays}
\end{center}
\end{figure}

If a compressional force along a specific direction is applied to a typical liquid droplet, it deforms in order to maintain its density and volume. The same occurs for quantum droplets made from quantum degenerate Bose mixtures. However, the binding mechanism for dipolar quantum droplets is anisotropic due to the anisotropic dipolar interaction. This anisotropy leads to a frustration of the dipolar quantum droplet when compressed along the magnetic field direction and the emergence of arrays of dipolar quantum droplets. Example images of such arrays are shown in \Fref{fig:arrays}\textbf{a} and \textbf{c} for different trap geometries.

The basic properties of dipolar quantum droplets in an external trap can be understood from a phase diagram calculated within a Gaussian variational ansatz \cite{Wachtler2016a, Bisset2016}. An example of such a phase diagram is shown in \Fref{fig:arrays}\textbf{b} for a system of dysprosium atoms confined in a cylindrically symmetric trapping potential. The regular BEC at large scattering lengths and the quantum droplet phase at low scattering lengths are connected via a continuous crossover for prolate and slightly oblate traps. A bistable region emerges with increasing strength of the confinement along the magnetic field direction. This bistable region features two local minima in the total energy per particle\textnew{, corresponding to the BEC and the quantum droplet solution} \cite{Wachtler2016a, Bisset2016}. The critical trap aspect ratio separates the continuous and the discontinuous phase transition. This critical aspect ratio was shown to depend on the total atom number and the mean trap frequency and has been measured in \cite{Ferrier-Barbut2018}.

In the initial droplet experiments \cite{Kadau2016, Ferrier-Barbut2016, Ferrier-Barbut2016a, Ferrier-Barbut2018}, multiple quantum droplets were observed after a quench of the scattering length in an oblate trap with $\lambda = \omega_\mathrm{z} / \omega_\mathrm{r} \approx 3$. Example images of droplet states with a different number of droplets are shown in \Fref{fig:arrays}\textbf{a}. These multi-droplet states appear upon crossing into the bistable region, due to the fragmentation of the system following a modulational instability. The multi-droplet states observed in the initial experiments are thus not the ground state of the system.

It is possible to make such a multi-droplet state the ground-state by confining the system in a cigar-shaped trap. This was first shown theoretically in \cite{Wenzel2017} and is illustrated in \Fref{fig:arrays}\textbf{d}. For large confinements along the magnetic field direction, the energy per particle of the two droplet state is lower than that of the single droplet. This behaviour is also observed in numerical simulations of the eGPE, where one finds states with an even larger number of droplets for stronger confinement. This behaviour was confirmed experimentally \cite{Wenzel2017}. Example images with different droplet numbers are shown in \Fref{fig:arrays}\textbf{c}. For large enough atom numbers, two-dimensional arrays of multiple quantum droplets eventually emerge as the ground state in cylindrically symmetric traps \cite{Baillie2018}.

\subsection{Collisions of droplets}

A powerful tool that can be used to probe the dynamical properties of self-bound objects are measurements of collisions between these objects. For example, such measurements have been performed with classical liquids \cite{Ashgriz1990, Qian1997, Pan2005}, helium clusters \cite{Vicente2000, Maris2003, Ishiguro2004} and atomic nuclei \cite{Andreyev2017, Bulgac2016, Magierski2017, Bulgac2017}. Measurements of collisions between quantum droplets thus promise to reveal important insights into the droplet properties.

\begin{figure}[t!]
\begin{center}
\begin{overpic}[width=0.95\textwidth]{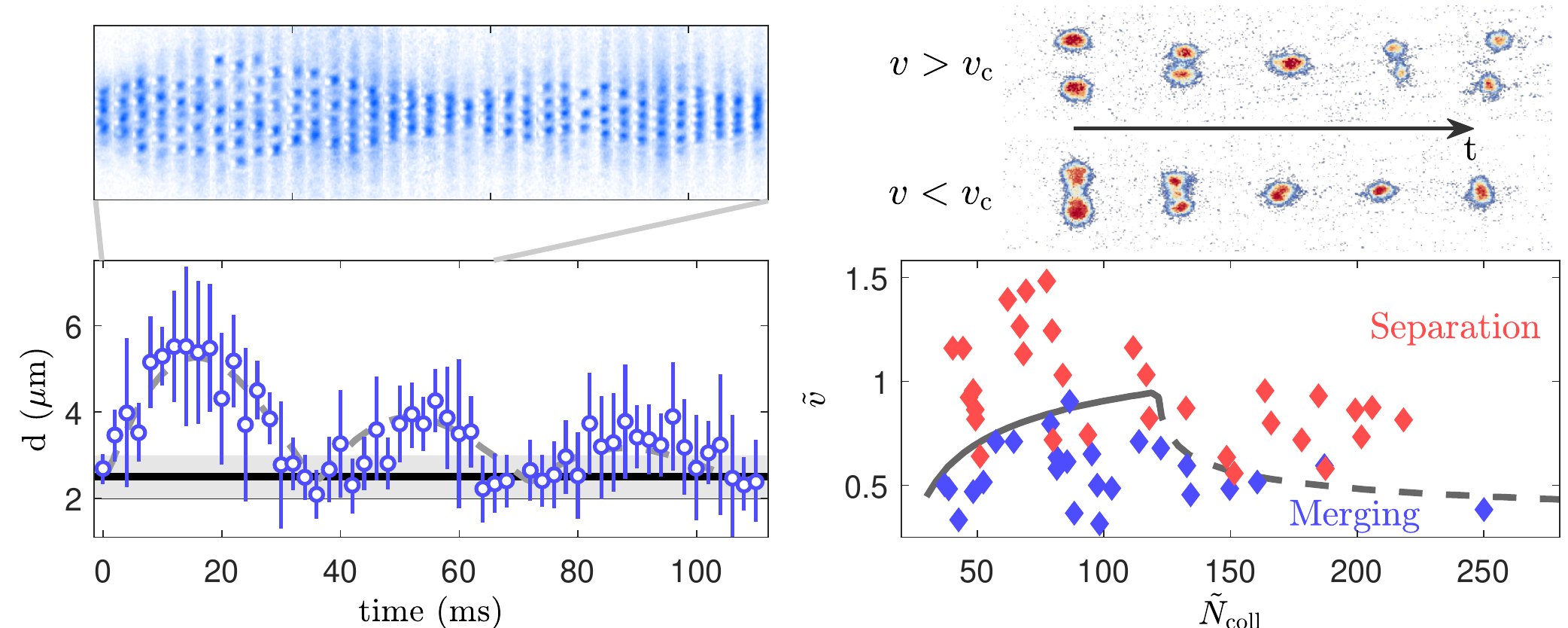}
\put(2, 37.5){\textbf{a}} 
\put(2, 21.5){\textbf{b}} 
\put(51.5, 37.5){\textbf{c}} 
\put(51.5, 28.5){\textbf{d}}
\put(51.5, 22){\textbf{e}}
\end{overpic}
\caption{
(\textbf{a}) Single-shot images of colliding dipolar quantum droplets for different evolution times. 
(\textbf{b}) Mean droplet distance $d$ as a function of the evolution time together with a guide to the eye representing a damped bouncing motion at the trap frequency (dashed gray line). 
(\textbf{c}), (\textbf{d}) Example images of collision measurements of independently created binary quantum droplets resulting in (\textbf{c}) a separation and (\textbf{d}) a merger of the droplets. (\textbf{e}) Outcomes of different droplet collision measurements as a function of the relative velocity $\tilde{v}$ and the atom number $\tilde{N}_\mathrm{coll}$ at the time of the collision, with the blue (red) diamonds indicating a merger (separation) of the droplets.
Adapted from \cite{Ferrier-Barbut2016a} and \cite{Ferioli2019}.
}
\label{fig:collision}
\end{center}
\end{figure}
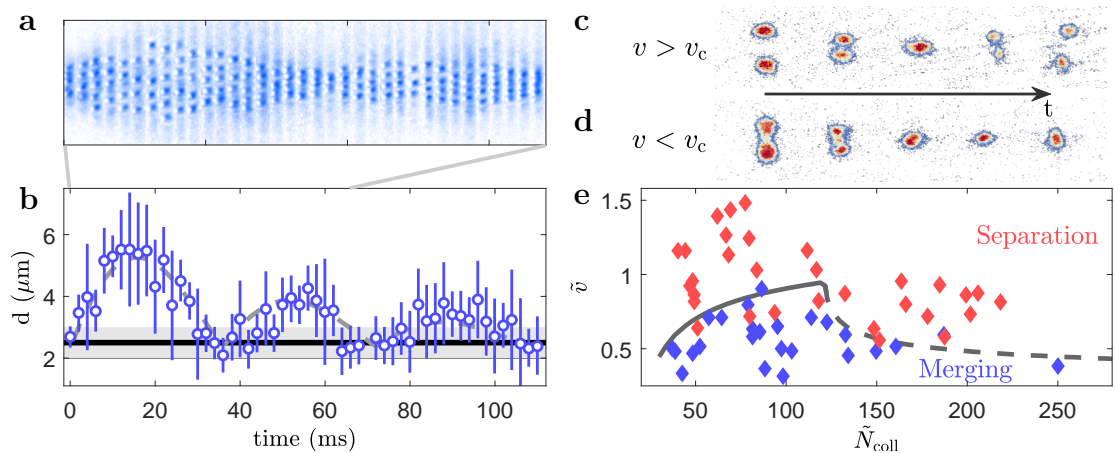

In the case of dipolar quantum droplets, collisions have been observed after loading a two-dimensional array into a one-dimensional waveguide with only weak confinement along one direction \cite{Ferrier-Barbut2016a}. Example images of several droplets at different evolution times in such a waveguide configuration are shown in \Fref{fig:collision}\textbf{a}. The droplets initially repel each other due to the repulsive dipole-dipole interaction. However, the weak confinement along the waveguide axis forces the droplets to reverse their motion and collide with each other. In \Fref{fig:collision}\textbf{b} the mean droplet distance is shown as a function of the evolution time, clearly showing that the droplets bounce off each other twice. The oscillation of the droplets in the waveguide is strongly damped because of the relative motion between droplets and the residual background atoms, and inelastic collisions. In the long-time limit, the droplets settle to an equilibrium distance of $d = \unit[2.5(5)]{\upmu m}$.

In Bose-Bose mixtures, the preparation of two quantum droplets requires spatially separated BECs as is typically realized with a double-well potential. In order for the droplets to collide the barrier separating them has to be removed. By further controlling the time at which the radial and vertical confinement is turned off, the velocity with which the self-bound droplets move towards one another can be precisely controlled \cite{Ferioli2019}. Similar to the collisions of classical liquids, two outcomes of the collision of quantum droplets are observed. For relative velocities smaller than a critical value $v_\mathrm{c}$, the droplets merge into a single droplet upon colliding. For larger velocities $v > v_\mathrm{c}$, the droplets separate again after collision and keep moving apart. These two outcomes are shown with example images in \Fref{fig:collision}\textbf{c} and \textbf{d}, respectively. 

While the critical collision velocity $v_\mathrm{c}$ depends on the mean atom number $\tilde{N}_\mathrm{coll}$ of the two droplets, the actual dependence is very different for small and large droplets. This can be understood in a liquid-drop model  \cite{Ferioli2019}, in which the surface tension is the important energy scale for large droplets. However, for small droplets there is no distinction between bulk and surface and as such the relevant energy scale is the binding energy. The significant change in the dependence of the critical velocity can thus be seen as evidence of a crossover from a compressible quantum droplet at small atom numbers to \textold{an}\textnew{a nearly} incompressible droplet at large atom numbers. In analogy to experiments with helium droplets \cite{Vicente2000, Maris2003, Ishiguro2004}, the study of the coalescence dynamics at small collision velocities may reveal important insights into the superfluid properties of the quantum droplets in the near future.

\section{Dipolar supersolid}

In \Sref{ch:arrays} we have seen how the anisotropy of the dipole-dipole interaction leads to the emergence of arrays of dipolar quantum droplets in confined geometries. These self-organized droplet arrays break the continuous translational symmetry of the system, naturally leading to the question of supersolidity. \textnew{A supersolid is a paradoxical state of matter that combines the frictionless flow of a superfluid with the crystal-like periodic density modulation of a solid \cite{Boninsegni2012}. While the search for supersolidity was initially focussed on helium \cite{Chan2013}, it has since moved to ultracold atomic system, thanks to their highly tunable interactions. The first experimental evidence of supersolid properties was reported, for BECs coupled to external light fields, either exploiting cavity-mediated long-range interactions between the atoms \cite{Leonard2017, Leonard2017a} or spin-orbit coupling \cite{Li2017}. While in these systems indeed two continuous symmetries are broken, the periodicity of the density modulation is set by the external light field, and therefore allowing no propagating phonon modes.}

\textnew{Using dipolar quantum gases, we have shown above that the transition to droplet arrays happens in a self-organizing fashion and is based purely on the intrinsic interactions of the constituent atoms. Therefore, these droplet crytsals allow for propagating phonon modes, one of the most paradigmatic properties of a solid. However, in} \textold{In} order to prove the supersolid nature \textnew{of arrays of quantum droplets}, the coexistence of spatial order and superfluidity needs to be established. However, the initial experimental study showed that the realized states rapidly lose their global phase coherence \cite{Wenzel2017}. While each droplet itself is superfluid, the whole system is not.

It was unclear whether a supersolid state can be realized in arrays of dipolar quantum droplets until a theoretical investigation of an infinitely extended system showed the coexistence of superfluidity and spatial order for a narrow range of the scattering length close to the phase transition \cite{Roccuzzo2019}. Roughly at the same time, measurements done by the group of G. Modugno in Pisa showed indications of phase coherence in droplet arrays in a cigar-shaped trap geometry \cite{Tanzi2019}. In a study from our group in Stuttgart that combined theory and experiment, the existence and the dynamical accessibility of such a phase-coherent array of dipolar quantum droplets was shown soon thereafter \cite{Bottcher2019}. Again shortly afterwards, the universality of the phenomena was shown in the group of F. Ferlaino in Innsbruck by observing phase-coherent droplet arrays in erbium, as well as in a different isotope of dysprosium \cite{Chomaz2019}. Together, the experimental results of the groups in Pisa, Stuttgart and Innsbruck undoubtedly proved the coexistence of spatial order and global phase coherence. 

\begin{figure}[t!]
\begin{center}
\begin{overpic}[width=0.95\textwidth]{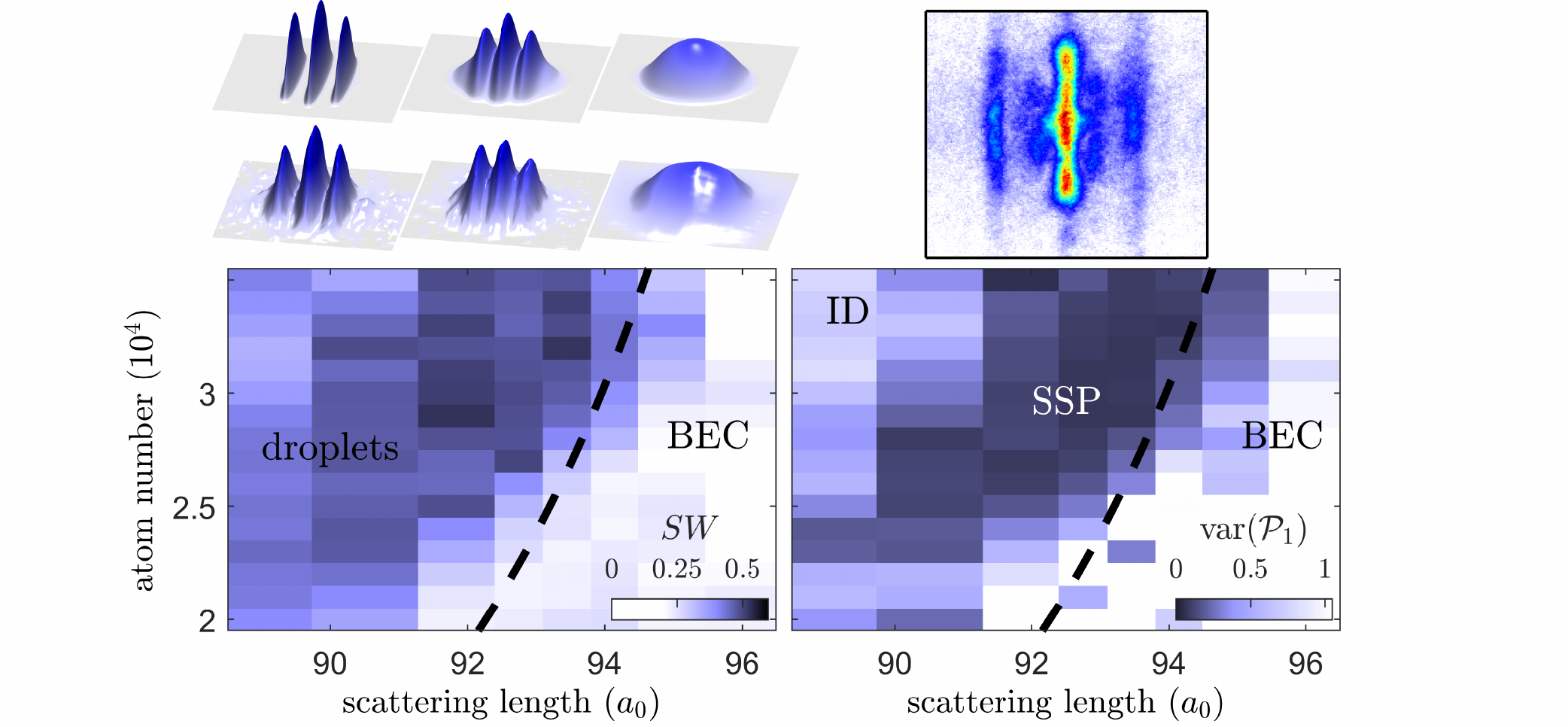}
\put(10, 44){\textbf{a}} 
\put(10, 30){\textbf{c}}
\put(54, 44){\textbf{b}} 
\put(54, 30.5){\textbf{d}}
\end{overpic}
\caption{
(\textbf{a}) Comparison of the theoretically (top) and experimentally (bottom) observed density profiles of an array of isolated droplets (left side), an array of quantum droplets immersed in a condensate background (center) and a regular BEC (right side) \cite{Bottcher2020}.
(\textbf{b}) Example image of the multi-wave interference in \cite{Bottcher2019}, showing the principal and minor interference fringes at a different spacing.
(\textbf{c}), (\textbf{d}) Experimental signatures of the phase diagram for (\textbf{c}) the in-situ density modulation and (\textbf{d}) the nearest-neighbour coherence. The strength of the density modulation in (\textbf{c}) is characterized by the spectral weight $SW$, which compares the contribution of Fourier amplitudes at finite momentum to the zero momentum contribution. 
(\textbf{d}) The Fourier transforms of the interference patterns reveal clear side peaks at the length scale corresponding to nearest- and next-nearest neighbours. Phase coherence leads to a well-reproducible interference pattern and thus a vanishing variance of the amplitude of these side peaks. The variance of the nearest-neighbour peak $\mathrm{var}(\mathcal{P}_1)$ allows to differentiate between three regimes -- isolated droplets (ID), phase-coherent droplets (supersolid phase, SSP) and a BEC.
The black dashed line in (\textbf{c}) and (\textbf{d}) indicates the theoretical phase boundary obtained from numerical simulations of the eGPE.
Adapted from \cite{Bottcher2019, Bottcher2020}.
}
\label{fig:supersolid1}
\end{center}
\end{figure}
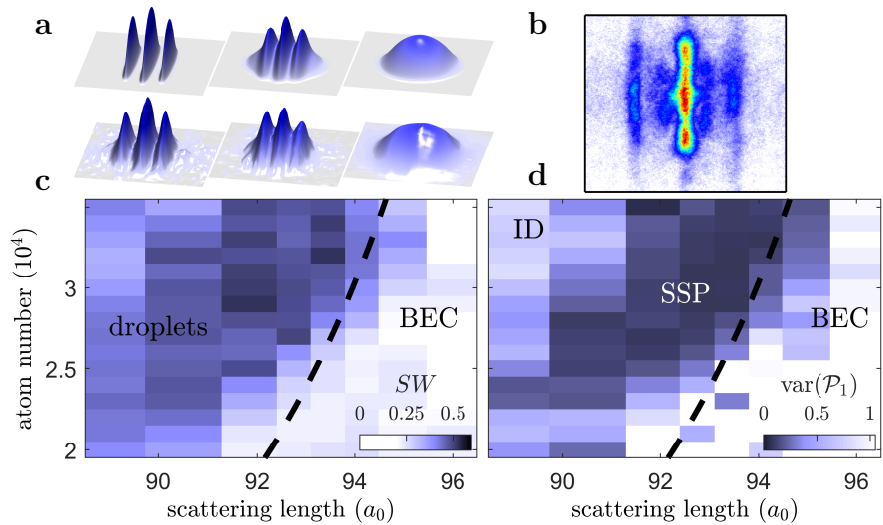

In \Fref{fig:supersolid1}\textbf{a}, we show the theoretically expected and the observed in-situ density profiles in the three different regimes -- an array of isolated quantum droplets, a droplet array immersed in a condensate background, and a regular BEC. %From the in-situ images, the strength of the density modulation can be directly characterized by looking at the spectral weight $SW$ at finite momenta and comparing it to the zero momentum contribution. The spectral weight as a function of the scattering length and the atom number is shown in \Fref{fig:supersolid1}\textbf{b} and allows us to distinguish between a regular BEC and an array of quantum droplets. 
The transition can be characterized by the strength of the density modulation, which is plotted in \Fref{fig:supersolid1}\textbf{c} as a function of the scattering length and atom number. 
The first step of proving superfluidity is to probe whether the system is phase-coherent. An example image of a time-of-flight interference measurement is shown in \Fref{fig:supersolid1}\textbf{b}. The observed interference patterns reveal evidence of the interference of multiple quantum droplets, allowing the characterization of the nearest- and next-nearest neighbour coherence \cite{Bottcher2019}. The experimental signature for the case of the nearest neighbour interference as shown in \Fref{fig:supersolid1}\textbf{d}, reveals three distinct regimes -- isolated and, hence, incoherent arrays of quantum droplets, phase-coherent droplet arrays and the regular BEC. 

\begin{figure}[t!]
\begin{center}
\begin{overpic}[width=0.95\textwidth]{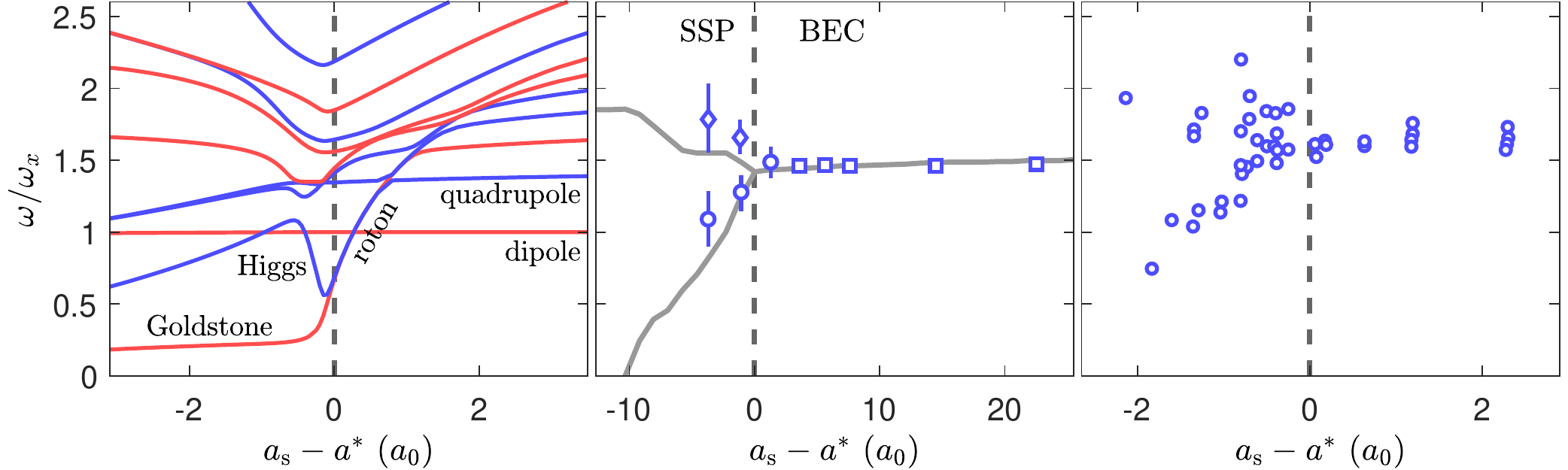}
\put(33, 7){\textbf{a}} 
\put(64.5, 7){\textbf{b}}
\put(95.5, 7){\textbf{c}} 
\end{overpic}
\caption{
(\textbf{a}) Calculated excitation frequencies $\omega / \omega_x$ of the lowest collective modes across the phase transition from a BEC to an array of quantum droplets in a cigar-shaped trap geometry with ${\omega_\mathrm{trap}=2\pi \,[30, 90, 110] \, \mathrm{Hz}}$. The blue (red) lines indicate an even (odd) parity of the density variation with respect to the weak trapping direction $\hat{x}$. 
(\textbf{b}), (\textbf{c}) Signature of the phase transition in the excitation energy of the axial quadrupole mode measured with (\textbf{b}) $^{162}$Dy and (\textbf{c}) $^{166}$Er. The measurements were done using trap frequencies of $2\pi \,[19 (2), 53(2), 81(2)] \, \mathrm{Hz}$ and $2\pi \,[259(2), 30(1), 170(1)] \, \mathrm{Hz}$ for the case of dysprosium and erbium, respectively.
Adapted from \cite{Hertkorn2019, Tanzi2019a, Natale2019}.
}
\label{fig:supersolid2}
\end{center}
\end{figure}

While the above measurements confirm the coexistence of spatial order and global phase coherence in a narrow range of the scattering length close to the phase transition, the presence of global phase coherence is not a sufficient criterion for superfluidity \cite{PitaevskiiStringariBook}. In order to prove the superfluid nature of the droplet arrays, the groups from Pisa, Stuttgart and Innsbruck studied the elementary excitations of the system. The breaking of a continuous symmetry at the superfluid-supersolid phase transition fundamentally affects the spectrum of collective excitations. The low-lying collective modes thus allow for fundamental insights into the symmetry breaking and the supersolid nature of the droplet arrays. In \Fref{fig:supersolid2}\textbf{a} we show the calculated energies of the low-lying collective modes as a function of the scattering length in a cigar-shaped trap geometry \cite{Hertkorn2019}.

For large scattering lengths, the typical collective mode spectrum of a dipolar BEC is obtained. The energetically lowest mode is the dipole mode which is completely decoupled from interactions and always features an excitation energy corresponding to the trap frequency $\omega_x$. Upon decreasing the scattering length, a softening of the roton mode is observed. The roton mode is characterized by a minimum in the dispersion relation at a finite momentum and corresponds to a perturbative density modulation on top of the condensate density distribution \cite{Santos2003, Giovanazzi2004, Ronen2007, Wilson2008, Jona-Lasinio2013, Natu2014, Shen2018, Corson2013, Chomaz2018, Petter2019}. In the considered trap geometry, the roton is comprised of two degenerate modes -- a symmetric and an anti-symmetric roton mode. As these roton modes soften, avoided crossings are observed between pairs of modes with equal parity. 

The softening of the roton triggers the phase transition to an array of quantum droplets, which in the considered finite system occurs at a finite excitation energy of the roton modes. The degeneracy of the even and odd roton modes is lifted with the emergence of a density modulation in the ground state. At smaller scattering lengths the excitation energy of the symmetric mode rapidly increases, whereas the excitation energy of the anti-symmetric mode further decreases. The symmetric mode features an oscillation between the droplet array and condensate background and can be understood as the Higgs amplitude excitation of the supersolid array of quantum droplets \cite{Hertkorn2019}. Close to the phase transition, the amplitude mode exists in an isolated state because of the energetic separation of the modes in the finite system. The amplitude mode hybridizes with the other symmetric modes as its excitation energy increases.

Close to the phase transition, collective modes with a larger excitation energy are affected by the rapidly increasing amplitude mode. The coupling of the higher symmetric modes with the amplitude mode seemingly leads to a bifurcation of the first quadrupole mode shortly after crossing the phase transition. The excitation energy of the first quadrupole mode across the phase transition is shown in \Fref{fig:supersolid2}\textbf{b} and \textbf{c} for the case of dysprosium \cite{Tanzi2019a} and erbium \cite{Natale2019}, respectively. An even higher-lying collective mode which is connected to superfluidity, is the scissors mode. Measurements of this particular mode show a decrease in the excitation energy upon crossing the phase transition. This can be understood as a stiffening due to the emerging density modulation \cite{Roccuzzo2019a, Tanzi2019b}.

\begin{figure}[t!]
\begin{center}
\begin{overpic}[width=0.95\textwidth]{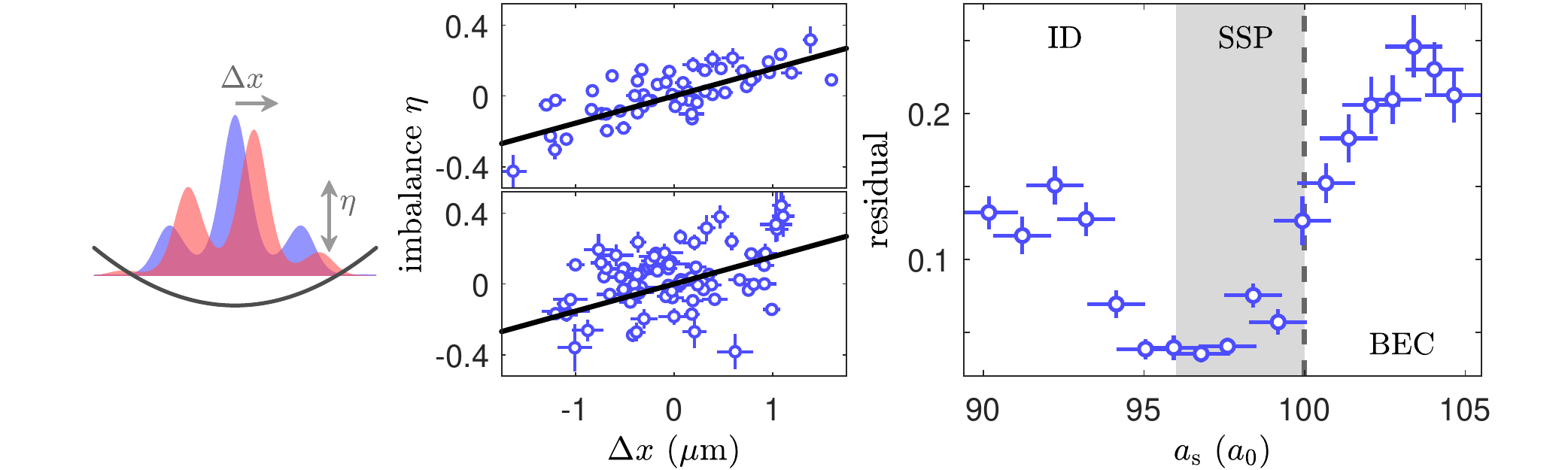}
\put(7, 24){\textbf{a}} 
\put(33, 26.5){\textbf{b}}
\put(33, 15){\textbf{c}}
\put(63, 26.5){\textbf{d}} 
\end{overpic}
\caption{
(\textbf{a}) Schematic of the low-energy Goldstone mode, featuring an out-of-phase and center of mass preserving oscillation between the droplet array and the superfluid background. A displacement of the droplet array by $\Delta x$ leads to an imbalance $\eta$ in the atom number of the side droplets.
(\textbf{b}), (\textbf{c}) Measured imbalances $\eta$ as a function of the measured displacements $\Delta x$ in (\textbf{b}) the supersolid region and (\textbf{c}) isolated droplet region. The black line shows the theoretically predicted correlation of the low-energy Goldstone mode.
(\textbf{d}) The low residual of the experimental data with respect to the theoretical prediction close to the phase transition clearly proves the existence of the low-energy Goldstone mode. The gray area indicates the phase-coherent region. 
Adapted from \cite{Guo2019, Bottcher2020}.
}
\label{fig:supersolid3}
\end{center}
\end{figure}
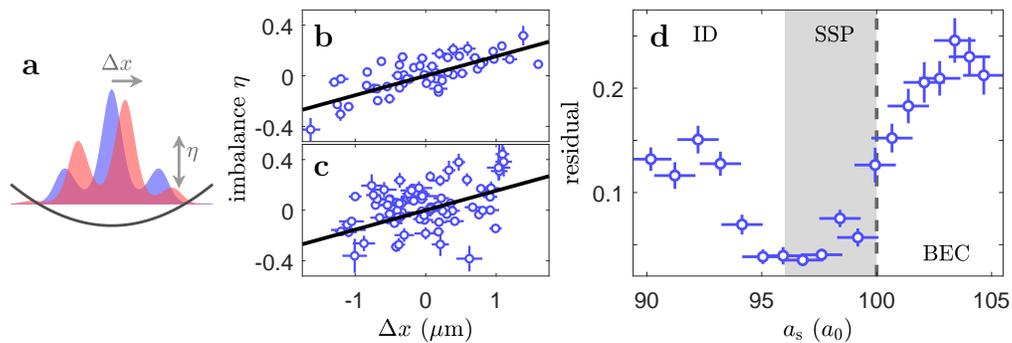

The collective mode most closely related to superfluidity is the low-energy Goldstone mode that emerges out of the anti-symmetric roton mode. This mode features an out-of-phase oscillation between the droplet array and the superfluid background, involving Josephson-like dynamics between the droplets and as a result highlights the superfluid nature of the state \cite{Guo2019}. \textnew{The existence of this mode is directly connected –- via the Goldstone theorem -- to the double symmetry breaking at the phase transition.} The dynamic of the low-energy Goldstone mode is illustrated in \Fref{fig:supersolid3}\textbf{a}. The precise interplay between crystal motion and superfluid counterflow preserves the center of mass and leads to a linear correlation between the array displacement $\Delta x$ and the imbalance $\eta$ between the side droplets. While the finite lifetime of the droplet array prevents the time-resolved measurement of the mode, its excitation leaves a trace on the spatial density distributions. This trace can be statistically mapped out by repeating the experiment many times, leading to the observed correlation between the imbalance and the displacement as shown in \Fref{fig:supersolid3}\textbf{b}. Lowering the scattering length and moving into the regime of isolated droplets, the correlation vanishes. Comparing the observed correlation to the theoretical prediction in \Fref{fig:supersolid3}\textbf{d} clearly proves the existence of the low-energy Goldstone mode close to the phase transition \cite{Guo2019}. All together these measurements of the collective modes confirm the genuine superfluidity, and thus supersolidity of the arrays of quantum droplets close to the phase transition.

Further studies of the supersolid droplet arrays include out-of-equilibrium dynamics \cite{Ilzhofer2019a} and high-energy Bragg spectroscopy \cite{Petter2020, Chomaz2020}. In the former, the supersolid is shown to re-establish a high-degree of phase coherence  on the timescale of the trap frequency, after a randomization of the phase of the individual droplets. The latter study shows the dynamical response of the dipolar supersolid to a two-photon Bragg excitation, which strongly reduces upon crossing the phase transition and finally vanishes for isolated droplets. \textnew{Lastly, measurements of density fluctuations across the superfluid-supersolid transition have been used to access the static structure factor, estimate the rotonic excitation spectrum and vizualize the corresponding symmetric and antisymmetric roton modes close to the critical point \cite{Hertkorn2020}. The observed strong thermal enhancement of the roton mode population in these measurements highlight the importance of finite temperature in the superfluid-supersolid phase transition.}

\section{Open questions and Outlook}

In the previous sections, we discussed measurements of the various characteristic properties of liquid quantum droplets and dipolar supersolids. The theoretical description based on the formalism first put forward by D. Petrov provides a satisfactory explanation for the properties and the existence of these new states of matter with fine-tuned interactions. However, precise measurements, for example of the critical atom number \cite{Bottcher2019a, Cabrera2018} or the roton softening in dipolar gases \cite{Petter2019}, have revealed discrepancies between theoretical predictions and experiments. In the following we focus on the current limits of our understanding of quantum droplets and dipolar supersolids, and briefly discuss ongoing experimental and theoretical efforts that aim to extend our knowledge and understanding of these new and peculiar states of matter.

The first point concerns the mean-field description of the eGPE, which  is fundamentally limited to the perturbative regime at small gas parameters \cite{Gautam2018}. Effects such as the quantum depletion and quantum correlations start to play a role at the large quantum droplet densities \cite{Bottcher2019a, Cikojevic2018}. Moreover, the beyond mean-field correction originating from quantum fluctuations is typically included in the description through a local density approximation. The validity of this approximation may not always be given in long-range interacting systems, close to a phase transition, in dense systems, and also for large dipolar strengths ($\edd > 1$). Furthermore, the mean-field collapse leading to the formation of the droplets is accompanied by imaginary Bogoliubov modes -- a soft phonon mode for the Bose-Bose mixtures and the roton mode for the dipolar system$^{\footnotemark}$\footnotetext{\textnew{Note that in finite-sized dipolar systems a set of local collapses occurs, which is typically referred to as a modulational instability \cite{Ferrier-Barbut2018b, Wilson2009, Parker2009}. This instability can only be directly associated with a softening roton mode, if $R_\mathrm{BEC} \times \left< k \right> \gg 1$. Here, $R_\mathrm{BEC}$ is the size of the condensate and $\left< k \right>$ the expectation value of a softening collective mode in the discrete spectrum of excitations.}} \cite{Ronen2006, Wilson2008, Wilson2009, Martin2012, Blakie2012, Bisset2013} -- which \textold{have to be set to zero artificially} \textnew{are neglected} \cite{Petrov2015, Hu2020, Ota2020}. While the presented theoretical description captures the underlying physics of quantum droplets, \textold{it is not self-consistent} \textnew{there is currently an extensive effort to improve it further}. 

\textold{To solve these problems and to find a self-consistent theoretical description of quantum droplets} \textnew{In this effort to improve the description of quantum droplets}, different theoretical approaches have been discussed, including the behaviour at the dimensional crossover \cite{Ilg2018, Edler2017, Zin2018}, \textnew{the diagrammatic Beliaev technique \cite{Gu2020}, the behaviour in one-dimensional optical lattices \cite{Morera2020, Morera2020a},} the hypernetted-chain Euler–Lagrange method \cite{Staudinger2018, Chakraborty1982, Campbell1972, Hebenstreit2016}, the Gaussian-state theory which includes squeezing effects \cite{Shi2019, Wang2020}, bosonic pairing \cite{Hu2020, Hu2020a} or the inclusion of higher order corrections to the Bogoliubov speed of sound \cite{Ota2020}. Complementary to these approaches, quantum Monte-Carlo simulations have been used to verify the formation of quantum droplets and to understand their properties \cite{Saito2016, Macia2016, Cinti2017, Cinti2017a, Bombin2017, Cikojevic2018, Cikojevic2019, Parisi2019}. These quantum Monte-Carlo methods intrinsically include particle correlations, quantum fluctuations, a finite system size, and a finite interaction range \cite{Cikojevic2019, Cikojevic2020}. However, these methods are limited to the usage of a simplified interaction potential because they cannot handle the bound molecular states in the complete interaction potential. 

Many theoretical works have focused on the investigation of quantum droplets in low-dimensional systems to circumvent these fundamental problems \cite{Petrov2016, Parisi2019, Parisi2020, Ilg2018, Edler2017}. In particular, they investigated one-dimensional systems where the energy functional of Bose-Bose mixtures does not suffer from the aforementioned imaginary Bogoliubov mode. Even in such a one-dimensional configuration, these quantum Monte-Carlo studies found small deviations from the predictions of the eGPE. This further shows that the theoretical description with the eGPE is not sufficient to quantitatively describe all physical properties of the quantum droplets to the highest precision. In order to solve these issues and to shed light on the underlying physics in the quantum droplet state, further and more precise measurements are necessary. As an example, in Ref. \cite{Bottcher2019a} it was shown that measurements of the critical atom number and the droplet density profile can be used as a sensitive benchmark for different theories in the near future.

Another fundamental question in the understanding of quantum droplets is the role played by finite temperatures. For quantum droplets in Bose-Bose mixtures, the proposed self-evaporation to zero temperature has so far not been experimentally verified. Furthermore, measurements of collective excitations are still lacking due to the finite lifetime in the experiments, but are necessary to understand the self-evaporation process. In dipolar quantum droplets, the unavoidable presence of thermal fluctuations at finite temperatures has been proposed to increase the critical atom number of a self-bound droplet, in particular for large scattering lengths and large atom numbers \cite{Boudjemaa2017, Aybar2018}. Moreover, the validity of the Born approximation for the dipolar scattering problem has been called into question, which would result in a temperature dependent systematic shift of the critical atom number to lower values \cite{Odziejewski2016, Odziejewski2017}. For this reason, an important goal in the experiments is the development of novel temperature probes that can be used to clarify and characterize the role played by a finite temperature. An example of such a probe was proposed in \cite{Wenzel2018a} and is based on spectroscopic measurements of embedded fermionic impurities, similar to measurements with liquid helium droplets \cite{Toennies2004}.

%All these fundamental questions of course also affect the formation of arrays of dipolar quantum droplets, and thus the understanding of the recently observed dipolar supersolid. 
In addition to the discussed challenges present in the understanding of quantum droplets, even more fundamental questions arise in the context of the dipolar supersolid. How general is the emergence of a supersolid state with respect to variations in the experimental parameters? How does finite temperature affect the supersolid phase transition? Upon evaporating into the supersolid state, are both symmetries broken simultaneously or sequentially? Do supersolid states exist in other systems, such as polar molecules \cite{Baranov2012, Capogrosso2010, Micheli2006, Pollet2010, Trefzger2009}? 

Moreover, in analogy to the situation in a Mott-insulator with perfect atom number correlations between different lattic sites \cite{Greiner2002}, the eGPE as a mean-field theory should also not be the correct description for arrays of isolated droplets with no wave function overlap \cite{Hertkorn2019}. The exact nature of the supersolid-to-isolated-droplet-array transition therefore remains to be explored in detail theoretically. 

With all these open questions, future studies of quantum droplets and dipolar supersolids promise even more exciting new discoveries and insights into the physics at play.

\ack
The authors would like to thank all previous members of the dysprosium team in Stuttgart, in particular I. Ferrier-Barbut, T. Maier, H. Blessing (Kadau), M. Schmitt and M. Wenzel. We further acknowledge many stimulating discussions over several years with H.P. Büchler, K. Jachymski, D. Petrov, F. Ferlaino, G. Modugno, A. Pelster, L. Santos, L. Tarruell, T. Pohl, S. Stringari, J. Boronat, F. Mazzanti, as well as with our colleagues within DFG Forschergruppe 2247 and the QUANT:ERA collaborative project MAQS. This work is supported by the German Research Foundation (DFG) within FOR2247 under Pf381/16-1, Pf381/20-1, and FUGG INST41/1056-1.

\section*{References}
\bibliographystyle{iopart-num} 
\bibliography{Bibliography}

\end{document}